\documentstyle[11pt,twoside,paspconf]{article}

\def\etal{{\it et al.\/\ }}
\def\ga{\lower 2pt \hbox{$\, \buildrel {\scriptstyle >}\over{\scriptstyle \sim}\
,$}}
\def\la{\lower 2pt \hbox{$\, \buildrel {\scriptstyle <}\over{\scriptstyle \sim}\
,$}}

\begin{document}

\title{VATT/Columbia Microlensing Survey of M31 and the Galaxy}

\author{Arlin P. S. Crotts$^1$}

\affil{Department of Astronomy, Columbia University, New York, NY~~10027}

\affil{\bigskip
$^1$ Guest observer, Vatican Advanced Technology Telescope (Alice
P.~Lennon Telescope \& Thomas J.~Bannan Astrophysical Facility),
Mt.~Graham, Arizona - and - 
Guest observer at Kitt Peak National Observatory (KPNO), NOAO, operated by
AURA, Inc.~for the NSF}

\begin{abstract}

We describe the results and outline the methods used in a search for
microlensing events affecting stars in the outer bulge and inner disk of M31,
due both to masses in M31 and the Galaxy.
These observations, from 1994 and 1995 on the Vatican Advanced Technology
Telescope and KPNO 4m, rule out masses over much of the range from
$\sim 10^{-7}~M_\odot$ to 0.08~$M_\odot$ as the primary constituents of the
mass of M31 and the Galaxy towards this field.
Furthermore we find six candidate events consistent with microlensing due to
masses of about $1~M_\odot$, but we suspect that some of these may
be cases where long-period red supergiant variables may be mistaken for
microlensing events.
Coverage from anticipated data should be helpful in determining if these
sources maintain a constant baseline, and therefore are best described by
microlensing events.
We analyzed our data using the new technique of ``difference image photometry''
(also called ``pixel lensing'').
A brief overview of this technique is included in the Appendix I.
This contribution summarizes two other recent papers (Tomaney \& Crotts 1996,
Crotts \& Tomaney 1996).

\end{abstract}

\section{Introduction}

The least radical candidate for galaxian dark matter is baryonic objects 
too large to be detected as dust or gas.
Objects of primordial composition and more massive than about $10^{-7}
M_{\odot}$ might be expected to resist evaporation until the present day (de
R\'ujula et al.~1992), while masses smaller than about $0.077 ~ M_\odot$ would
fail to ignite as stars (Burrows et al.~1993).
Gravitational microlensing might reveal individual objects via their effects on
background stars (Paczynski 1986).
Such searches have recently taken place towards the Large Magellanic Cloud
(LMC) and Bulge, with searches towards the LMC ruling out most of the dark
matter being composed of substellar-mass objects (Aubourg et al.~1995, Alcock
et al.~1996) heavier than about $10^{-6} M_\odot$, while suggesting that much
dark matter has the same component mass as low-mass stars (Alcock et al.~1996).
Given the uncertainty of the Galactic halo's distribution of MAssive, Compact,
Halo Objects (MACHOs) and therefore the lensing geometry involved, the relation
of mass to observed microlensing timescale is still unclear.

In part because of its unique geometry with respect to Earth and partially due
to high predicted lensing optical depths ($\tau$), M31 is a uniquely powerful
venue for studying microlensing.
Early we realized that an M31 microlensing survey would show many advantages if
studying such a distant, crowded field of stars could be made practical.
Such an approach is outlined by Crotts (1992, hereafter C92) with a complete
description found in Tomaney and Crotts (1996, hereafter TC) and summarized in
Appendix I.
By subtracting images in a time sequence, then performing difference image
photometry, we can study the residual point sources due to variables, while the
signals from the many crowded, non-varying stars subtract away.
Thereby we exploit
advantages inherent in studying M31: 1) very small component mass limits, due
to the small angle subtended by the photosphere of M31 stars, 2) the ability to
survey different parts of M31, thereby studying the spatial distribution of
microlensing objects, 3) the ability to study many stars at once in fields of
high $\tau$, thereby detecting events in short periods of observation, and 4)
well-constrained microlensing geometry, due to the lensing mass being
concentrated over the center of the galaxy, allowing a better determination of
the MACHO mass.
We exploit this for several candidate events, described below and in Crotts
\& Tomaney 1996 (CT).

\section{VATT and KPNO Observations and Results}

We observed at the Vatican Advanced Technology Telescope (VATT) on 33 nights
from Dec 1994 to Dec 1995, and with the KPNO 4m PFCCD on 24-27 Sep 1994 and
28 Aug 1995.
The field looks past the bulge, intercepting the far side of the disk
along the minor axis.
Assuming a distance to M31 of 770~kpc, the VATT field covers 0.4 to 2.9~kpc
along the minor axis, which projects to 1.8 to 13.1~kpc along the disk,
assuming a $77^\circ$ inclination.
Typically our M31 sample stars are red giants (RGs).
We choose two bands to maximize the number of photons from RGs: non-standard R
and I, slightly broader than their conventional equivalents: R extends from
$\lambda5700$ (just beyond the [O~I]~$\lambda5577$ night sky line) to
$\lambda7100$, and filter I extends from $\lambda7300$ to $\lambda10300$
(5\% power points).
The Sep 1994 KPNO data also constrain short-timescale microlensing.
The KPNO analysis is based on coadded exposures totaling 12.5~min in
both R and I.
Integrating the {\it unsaturated} regions beyond the central few arcmin$^2$ of
the bulge yields a mean surface brightness of galaxy plus sky of $\mu_R =19.35$
and $\mu_I = 18.30$.
This corresponds to a mean coadded $\langle S/N \rangle=16$ for an R = 22.5
star.
Brighter than this cutoff we are sensitive to $6.7\times10^5$ stars at greater
$\langle S/N \rangle$ in this single field (using our number density estimation
method in Appendix II).
With CCD readout time, this corresponds to a time resolution of 50~min per
coadded frame in a given band (about four times the photospheric-radius
crossing time for a RG in M31).
Such time resolution yields sensitivity to Galactic MACHOs down to 
$\sim2\times10^{-7} M_\odot$.

The process of difference image photometry consists of careful flat-fielding,
coordinate registration and photometric scaling of the data, followed by point
spread function (PSF) matching between frames (detailed in the Appendix I).
(The error bars presented with the lightcurves in Figures 1 and 2 show the
fluctuations in the difference image on the scale of the PSF in regions
adjacent to each residual source.)~
{\bf Timescales longer than 1 day}: sources were catalogued by requiring at
least a 4$\sigma$ detection in at least
two nightly sums (or 6$\sigma$ in the 24\% of the image containing the bulge
and closest to the minor axis).
We locate over 2000 sources within the VATT field.
{\bf Short timescale (KPNO) analysis}: sources were originally identified by
eye in the difference frame at typical S/N ratio of 6 or greater.
Thus these S/N ratios correspond to $12\sigma_{photon}$ given our determined
technique limits in Appendix I.
For a fiducial minimum amplification of $>34\%$ corresponding to stars passing
within the Einstein ring of the lens (Paczynski 1986), we must therefore detect
the original star at $36\sigma_{photon}$ to detect this minimum amplification
at a S/N $> 6$ in the difference frame.
We require such a detection in the same place in at least two difference
frames.
{\bf Candidate events:} nightly sums reveal no source on only two consecutive
nights, and none, with one exception, consistent with microlensing events on
any but nearly the longest timescales sampled by our survey.
In the latter cases, we portray the lightcurves of the six candidate events in
Figure 1, and other information in Table 1, including their positions (J2000)
and distance along M31's minor axis ($d$).
Assuming that they are microlensing events, other parameters can also
be extracted e.g.~lensing impact parameter (normalized to the Einstein radius:
$u_o = u/R_e$).
Not given are other fit parameters, the duration (Einstein radius crossing time
$t_e$), time of peak amplification, source baseline magnitude, and flux
zero-point offset due to image subtraction.
Also we give the goodness of the best lensing fit (for point sources and
masses), and the most probable lens mass.
It appears that fit residuals are slightly larger than expected from
photometric measurement error alone, seen particularly as a surplus in the
number of 3$\sigma$ or greater residuals, which are inconsistent with
neighboring points.
We investigate these noise sources in TC and elsewhere (Tomaney et al., in
preparation).
We stress that {\it we do not claim that these are microlensing events at least
until their lightcurves are observed to fall and remain at the pre-event
baseline} during the 1996 observing season or thereafter.

\begin{table}
\caption{Candidate M31 Microlensing Events}
\label{tbl-1}
\begin{center}
\begin{tabular}{cccccc}
\tableline
RA      & Dec     & Minor Axis & Probable        & Impact     & $\chi^2/\nu$ \\
(J2000) & (J2000) & Distance   & Mass,           & Parameter  &              \\
        &         & $d$, (kpc) & $m$ ($M_\odot$) & vs.~$R_e$, $u_o$ &        \\
\tableline
0$^h42^m55^s$.7 & $+41^\circ14^\prime27^{\prime\prime}$ & 0.59 & (0.09)$^a$ & 0.648 & 0.75 \\
0$^h42^m42^s$.3 & $+41^\circ11^\prime~2^{\prime\prime}$ & 0.62 & (4.3)$^a$  & 0.369 & 1.67 \\
0$^h42^m54^s$.1 & $+41^\circ10^\prime55^{\prime\prime}$ & 1.02 &  0.90$^b$  & 0.680 & 1.35 \\
0$^h43^m14^s$.8 & $+41^\circ12^\prime32^{\prime\prime}$ & 1.49 &  0.90$^b$  & 0.501 & 1.62 \\
0$^h43^m22^s$.6 & $+41^\circ~5^\prime52^{\prime\prime}$ & 2.67 &  0.32      & 0.590 & 2.23 \\
0$^h43^m49^s$.0 & $+41^\circ11^\prime28^{\prime\prime}$ & 2.77 &  0.49      & 0.690 & 2.01 \\
\tableline
\multicolumn{4}{l}{$^a$Source probably in bulge, so mass estimate is unreliable.} & & \\
\multicolumn{4}{l}{$^b$Source in bulge or disk; mass estimate assumes disk.} & & \\
\end{tabular}
\end{center}
\end{table}

\begin{figure}
\vspace{0.70in}
\plotfiddle{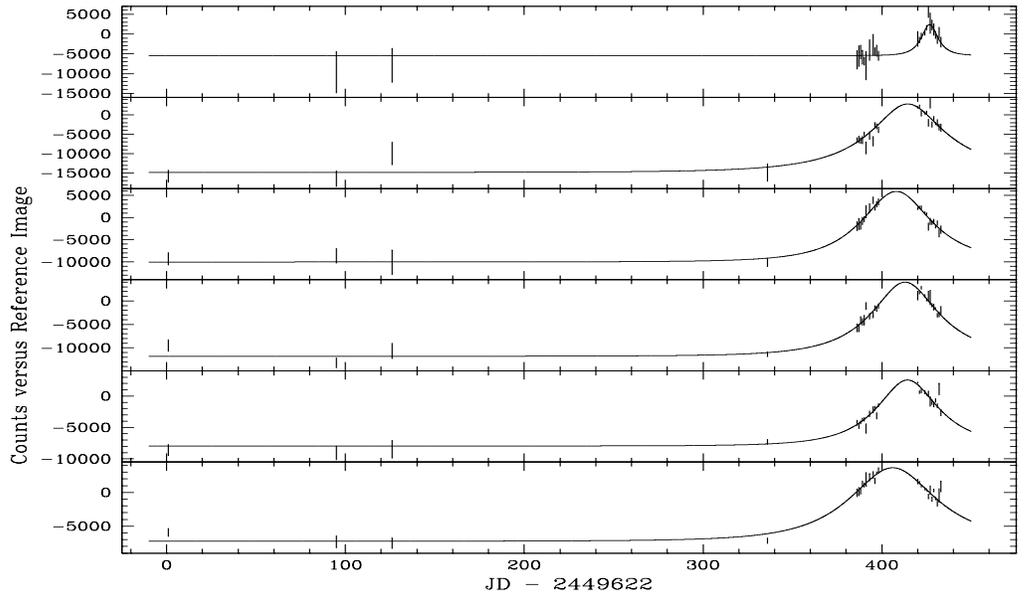}{2.0in}{0.}{70.}{50.}{-200}{-130}
\caption{
Lightcurves of M31 microlensing candidates from Table 1
}
\end{figure}

\begin{figure}
\vspace{-1.70in}
\plotfiddle{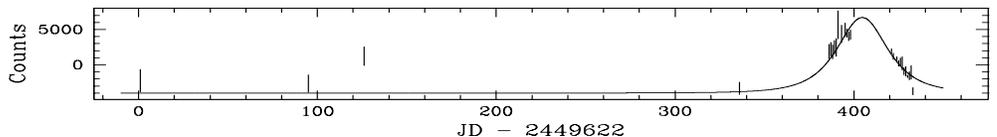}{2.0in}{0.}{70.}{50.}{-200}{-200}
\caption{
Lightcurve of ``mira'' at $0^h43^m37^s.9~+41^\circ14^\prime
57^{\prime\prime}$ (J2000)
}
\end{figure}

A reason for caution is the variable star lightcurve shown in Figure 2.
It is fit well by a microlensing lightcurve during its rise and fall, but does
not maintain a consistent baseline before and after the event.
Upon inspection of lightcurves of Mira-type variables (Wesselink 1987), we find
a small fraction whose maximum light behavior mimics microlensing lightcurves.
Further reasons for suspicion is the similarity in timescales to those of miras
(except the first event), and similar shapes, indicated by $u_o$ values which
cluster around 0.6 (except for the second event).
Additionally, it is strange that all sources have $R\approx21$, close to the
magnitude that would correspond to a mira pulsation (given the inferred $u_o$),
but brighter than what we might expect for lensed sources given the luminosity
function of stars in the field.
The reality of these events can be tested in terms of the distribution of $u_o$
values via a one-sample Kolmogorov-Smirnov (K-S) test.
The theoretical distribution is derived using a luminosity function in $R$
$\phi \propto 10^{\alpha R}$, where $\alpha = 0.59$
(see Appendix II).
The largest value of the K-S distance $D$ occurs at the smallest observed $u_o=
0.369$, due to the lack of small $u_o$ events, and has a value $D\approx0.7$.
Assuming that all six candidates are true microlensing events, the null
hypothesis (consistency with microlensing) is rejected at the 99.5\% level.
If half the candidates are microlensing events (and the minimum still
$u_o=0.369$), the null hypothesis is rejected at approximately the 90\% level.
It is unlikely that all of the events are due to microlensing, but this test
cannot rule out that a large fraction may be.
These caveats aside, {\it if}~these are microlensing events, we conclude
the following:
the first and second events land in the bulge-dominated region, and hence 
likely involve bulge sources.
In the case of third through sixth events, the most probable source-lens
distance is $d/cos~i$, allowing us to compute a most likely mass (see Table 1).

\section{Discussion}

Several approaches have been taken to estimating the predicted $\tau$ in M31
due to its own mass distribution.
Initially C92 just approximated the entire mass of M31 as an $r^{_{-2}}$
density distribution, which produces an optical depth for far-side disk stars
of $\tau \approx 10^{-5}$.
Several approaches have been taken since then (Jetzer 1994, Han \& Gould 1996),
and M31 disk (Gould 1994) and Galactic halo (Paczynski 1986) add lesser
amounts.
When all components are summed together at least $\tau\approx5\times10^{-6}$
results throughout the field, which is the value that we will adopt for the
sake of discussion.
From our constraints on the luminosity function of stars in our
field (TC and Appendix II), we have estimate that we are sensitive to
detectable microlensing of any of $6.9\times10^5$ stars in our VATT field.
These data are primarily sensitive to timescales ranging from 2$^d$ to 10$^d$,
corresponding to 0.003~$M_\odot$ to 0.08~$M_\odot$.
The predicted number of events for this mass range given a $\tau_{Gal+M31}$
of $5\times 10^{-6}$ is 45 to 7 events.
Except for one possible detection at the upper end of this range, we find
no events on these timescales, thereby eliminating this mass range as a 100\%
contribution to the mass of M31 at considerably better than 95\% confidence.
On the other hand, we expect to detect approximately 2 events (given 100\%
efficiency) if the mass of M31 is made entirely of $1~M_\odot$ objects, while
we see six candidates, half of which are at this scale or larger.
This argues that some of these may not be caused by microlensing.
The search of the KPNO data yielded 139 detected sources over the four nights
of data.
None of these are consistent with microlensing events.
Requiring multiple exposures per event in the KPNO data, in both bands, we have
19 independent 50$^m$ sample times, corresponding to a mass scale of about
$2\times 10^{-7} {\rm M}_\odot$.
Thus at this timescale we are sensitive to $7.9\times10^6$ star-epochs (see
Appendix II), which corresponds to a $2\sigma$ optical depth of $\tau < 5\times
10^{-7}$, for a $\delta$-function in $m$.
Over 8$^h$ timescales corresponding to masses of $8\times 10^{-5} {\rm
M}_\odot$ we have two sample times corresponding to the two last nights, which
corresponds to a $2\sigma$ limit of $\tau < 3.3\times 10^{-6}$ applying both to
M31 and Galactic halo MACHOs.
Given a simple spherical Galactic halo estimate towards M31 of $\tau =
1.0\times10^{-6}$ (Paczynski 1986), we eliminate the possibility at the 2-sigma
confidence that the Galactic halo is comprised of a single mass population of
MACHOs in the Earth mass range (${\rm M}_\oplus = 3\times 10^{-6}
{\rm M}_\odot$).
For M31 plus Galactic MACHOs, we can expect values for $\tau$ of 5-10$\times
10^{-6}$, which is a factor of several times larger than our 2-sigma limit
quoted above for $8\times 10^{-5} {\rm M}_\odot$.
Despite the systematic uncertainties involved, it would appear that these mass
ranges are ruled out for a 100\% contribution to the both dark matter MACHO 
halos.

\section {Appendix I: Difference Image Photometry}

To cope with severely crowded conditions unavoidable in ground-based imaging of
fields, C92 suggested registering a sequence of CCD images to common
coordinates, scaling to the same photometric intensity and subtracting images
from a high S/N (signal to noise) template image.
Since over some timescales of interest for microlensing events (10~min up to a
few months), {\it most} stars will not be varying (at least above some given
detection limit in flux change), then those stars that do vary over the time
span between the test frame and the reference frame may be detected in the
difference frame as isolated positive or negative point sources.
We call this Difference Image Photometry (DIP) (also dubbed ``pixel lensing'' -
Gould 1996). 
We match the PSF between frames as follows:
first, ignoring noise, consider a frame $r$ with a narrower PSF than a frame
$i$, $ i = r * k$, where $k$ is a convolution kernel that describes the
difference in the seeing and guiding between the two frames.
The Fourier transform (FT) of these three variables has the form, $ FT (i) =
FT (r) \times FT (k)$, then, $ k = FT [ FT (i) / FT (r) ]$.
Thus $k$ can be determined empirically with a high S/N isolated star on a frame
pair.
Convolving the good seeing frame with this kernel will provide a match to the
PSF of the poorer seeing frame.
In practice the determination of $k$ is not straightforward since the high
frequency fourier components are dominated by the noise in the PSF wings,
where the signal is weakest.
An effective method of dealing with this problem was determined by Ciardullo,
Tamblyn \& Phillips (1990).
Since the FT of a typical PSF is roughly Gaussian, the convolution kernel will
also be approximately Gaussian.
By modeling the high S/N low-frequency components of the PSF FT with an
elliptical Gaussian, these noise-contaminated components can be replaced with
the model fit yielding a nearly ideal convolution kernel.
We used IRAF DAOPHOT to perform photometry of the sources on the difference
frames.
The intrinsic accuracy of the photometric scaling accounts for
$1.4\sigma_{photon}$ seeing element residuals in the difference frame.
RR Lyraes and other ``noise'' raise this to about $2\sigma_{photon}$.

The KPNO 4m PFCCD has a highly spatially variable PSF, requiring us to map and
interpolate the convolution kernel over the image (explored at length in TC).
Reducing the spatial dependence of the PSF correspondingly reduces the effect
of changes in telescope focus inducing a spatial dependence in the kernel.
In the case of our VATT survey, we installed a doublet biconvex achromat,
designed with the aid of Richard Buchroeder of Tucson, Arizona, that produces
uniform 20 micron ($0.25''$) diameter spots and best focus over the entire CCD.
This resulted in $1-4$ PSF kernels being required for the VATT, versus
approximately 200, together with the kernal interpolation algorithm, for the
KPNO 4m PFCCD data, which covers only 2.1 times the solid angle in field of
view.

To highlight the sensitivity of the image differencing technique Figure 3 shows
two consecutive combined $36''\times 36''$ R band subimages separated by $50^m$
in time taken on the last night.
The upper panels are the undifferenced images, but with the large scale median
smoothed galaxy and sky background subtracted.
The lower panels are the result of differencing the original images with a
reference image comprising an average of all images taken on the previous
night.
The difference image for the second frame shows a clear detection ($20\sigma$)
of variability over the previous frame.
Remarkably, the eye cannot discern any indication of such variability in the
raw frames.

\begin{figure}
\vspace{0.70in}
\plotfiddle{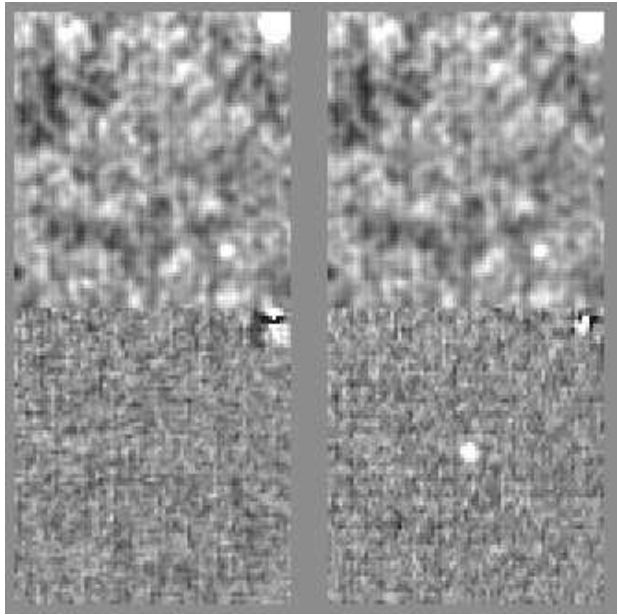}{2.0in}{0.}{40.}{40.}{-450}{-850}
\caption{
Illustration of difference image photometry (see text).
}
\end{figure}

\vspace{-0.1in}
\section{Appendix II: Luminosity Function and Source Sample}
 
(See TC for a more detailed treatment of the following issues.)
The number of stars per pixel detectable above a certain $S/N$ threshold is
crucial in understanding the conversion of an event rate to an optical depth
due to gravitational lensing.
In turn, the number of stars per pixel depends on the shape of the luminosity
function, and its first-moment integral, the surface brightness.
Calibrating our surface brightness data is straightforward, despite the lack of
a night sky brightness determination, in that surface brightness photometry is
published for our field (Walterbos \& Kennicutt 1987).
We have developed a simple technique for recovering the shape of the luminosity
function, based on the distribution of local surface brightness in various
pixels.
Figure 4 shows the number of pixels in a small subframe of our image 
plotted as a histogram versus the signal per pixel in ADU, once the 
sky background has been subtracted.
The thick solid line shows the plot for our actual data, for a subframe in
which the spatial gradient of counts has been removed without affecting the
mean count per pixel.
The other curves denote pixel histograms for several cases of simulated star
fields composed from luminosity functions of the form $\phi(L) dL \propto
L^{-0.4\alpha} dL \propto 10^{\alpha m} dm$, where $m$ is apparent magnitude
(in which $\alpha=0.60$ for the thin solid line, 0.55 for the dotted.)
The normalization of the simulated histograms is set only by requiring 
that the first moment of the distribution, the surface brightness, is 
maintained.
We note that higher moments are also recovered simultaneously, allowing the
entire shape of the real data's histogram to be recovered with the proper
choice of $\alpha$, in this case $\alpha = 0.59 \pm 0.01$.
This value of $\alpha$ is similar to those found for bright stars in comparable
environments (TC and references therein).
The power law approximation must break down at the faint end of this
distribution, however, given the presence of the horizontal branch at
$R=25.0$.
 
Usually in estimating event rates one selects a stellar sample above a
threshold apparent brightness, and a minimum amplification, which implies a
set lensing cross-section per MACHO mass given known Observer-Lens-Source
distances.
The product of the implied optical depth corresponding to the number density of
MACHOs and the number of stars in the sample produces the mean number of
lensing events at a given time, and the number of events over a period of
observation (much longer than the event) is then inversely proportional to the
typical event duration.
However, for a lensed star sample surveyed by DIP, we cannot impose stellar
brightness nor amplification thresholds, but instead a minimum flux change.
This effectively increases the number of events per star brighter than a given
threshold by about a factor of three (TC).

\begin{figure}
\vspace{0.90in}
\plotfiddle{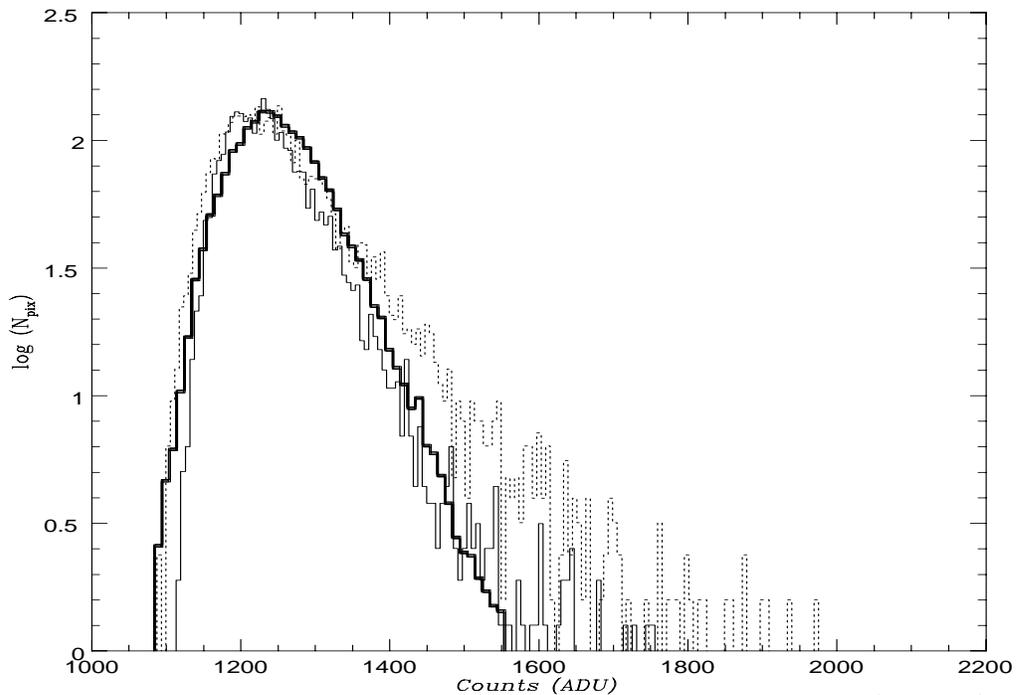}{2.0in}{0.}{70.}{50.}{-220}{-100}
\caption{
Pixel histograms for model luminosity functions (see text)
}
\end{figure}


\vspace{-0.15in}
\begin{references}
\vspace{-0.05in}
 
\reference{} Alcock, C. \etal 1996, preprint

\reference{} Aubourg, E. \etal 1995, A\&A, 301, 1

\reference{} Burrows, A., Hubbard, W.B., Saumon, D.~\& Lunine, J.I.~1993, ApJ,
406, 158

\reference{} Ciardullo, R., Tamblyn, P., \& Phillips, A.C.~1990, PASP, 102, 1113

\reference{} Crotts, A.P.S.~1992 ApJ, 399, L43 (C92)

\reference{} Crotts, A.P.S.~\& Tomaney, A.~1996, ApJ Letters, in press (CT)
 
\reference{} de R\'ujula, A., Jetzer, Ph.~\& Mass\'o, E.~1992, A\&A, 254, 99

\reference{} Gould, A.~1994, ApJ, 435, 573

\reference{} Gould, A.~1996, preprint

\reference{} Han, C.~\& Gould, A.~1996, preprint

\reference{} Hodge, P., Lee, M.G., \& Mateo, M.~1988 ApJ, 324, 172

\reference{} Jetzer, Ph.~1994, A\&A, 286, 426

\reference{} Paczynski, B.~1986, ApJ, 304, 1

\reference{} Tomaney, A.~\& Crotts, A.P.S.~1996, AJ, in press (TC)

\reference{} Walterbos, R.A.M.~\& Kennicutt, R.C.~1987, A\&AS, 69, 311
 
\reference{} Wesselink, T.~1987, PhD thesis (Nijmegen)

\end{references}
\end{document}